# Scaling properties in the production range of shear dominated flows

C. M. Casciola, P. Gualtieri, B. Jacob, and R. Piva

*Dipartimento di Meccanica e Aeronautica, Università di Roma* La Sapienza*, Via Eudossiana 18, 00184 Roma, ITALY*

(Dated: February 16, 2005)

Recent developments in turbulence are focused on the effect of large scale anisotropy on the small scale statistics of velocity increments. According to Kolmogorov, isotropy is recovered in the large Reynolds number limit as the scale is reduced and, in the so-called inertial range, universal features - namely the scaling exponents of structure functions - emerge clearly. However this picture is violated in a number of cases, typically in the high shear region of wall bounded flows. The common opinion ascribes this effect to the contamination of the inertial range by the larger anisotropic scales, i.e. the residual anisotropy is assumed as a weak perturbation of an otherwise isotropic dynamics. In this case, given the rotational invariance of the Navier-Stokes equations, the isotropic component of the structure functions keeps the same exponents of isotropic turbulence. This kind of reasoning fails when the anisotropic effects are strong as in the production range of shear dominated flows. This regime is analyzed here by means of both numerical and experimental data for a homogeneous shear flow. A well defined scaling behavior is found to exist, with exponents which differ substantially from those of classical isotropic turbulence. Contrary to what predicted by the perturbation approach, such a deep alteration concerns the isotropic sector itself. The general validity of these results is discussed in the context of turbulence near solid walls, where more appropriate closure models for the coarse grained Navier-Stokes equations would be advisable.

## I. Introduction.

One of the issues in fluid dynamics concerns the small scale behavior of turbulence. According to Kolmogorov, turbulent flows at sufficiently large Reynolds number develop a range of scales, much smaller than the integral scale of the system, where the dynamics is isotropic, universal and independent of viscosity. In these conditions statistical observables such as the moments of the velocity increments between two points are expected in the form of power laws of separation with universal exponents. These scaling laws are normally exploited in closures for the Navier-Stokes equations, coarse grained at inertial scales. In many cases, however, inertial range predictions are violated and this entails the failure of the related closures. It typically occurs in shear dominated flows, such as boundary layers near solid walls.

Actually, shear flows exhibit distinct ranges of scales, see Fig. 1 for nomenclature and definitions. When the Reynolds number is sufficiently large, a classical inertial subrange [1] sets in between the dissipative scale, $\eta$, and the energy injection scale, $\ell$. In equilibrium shear flows, where production and dissipation balance, $\ell = L_s = \sqrt{\epsilon/S^3}$. In the inertial subrange the longitudinal spectrum behaves like $E_{xx}(k_x) \propto \epsilon^{2/3} k_x^{-5/3}$, and the longitudinal structure functions exhibit scaling laws, $S_n \propto r_x^{\zeta(n)}$ [2], with exponents consistent with homogeneous isotropic data. There is however a residual effect of the anisotropy, as shown by the $n^{th}$ order structure tensors $S_{\alpha_1,...,\alpha_n}^{(n)}(r_\beta) = \langle \delta u_{\alpha_1} \ldots \delta u_{\alpha_n} \rangle$, where $\delta u_\alpha = u_\alpha(x_\beta + r_\beta) - u_\alpha(x_\beta)$ [3]. The proper tool to address this issue is by projecting the relevant tensors on the invariant subspaces of the rotation group, the so-called SO(3) decomposition. Due to the rotational invariance of the Navier-Stokes equations, such SO(3) components are found to manifest pure scaling laws in terms of separation $r = \sqrt{r_\alpha r_\alpha}$. The exponents are independent of the flow details and the isotropic sector behaves exactly as in isotropic turbulence. In the inertial subrange, the anisotropic contributions are subleading and the isotropic sector controls the small scale asymptotics [3–5]. All these results rest on the assumption that, at inertial scales, anisotropy is at most a weak perturbation of an otherwise isotropic dynamics.

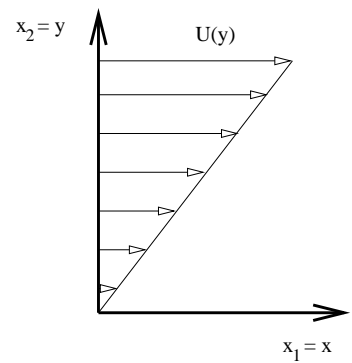

FIG. 1: Sketch of the shear flow and nomenclature: the mean flow $U(y)$, in the $x \equiv x_1$ direction, is a function of $y \equiv x_2$, with $z \equiv x_3$. For a linear mean profile, the shear rate $S = dU/dy$ is constant and the flow is spatially homogeneous. In wall bounded flows the shear rate depends on distance from the wall. The field is the sum of average and fluctuation, with the latter denoted by $u_\alpha, \alpha = 1, 2, 3$ - i.e. $u_{x/y/z}$, or $u, v, w$. The production of turbulent kinetic energy is $\Pi = S \langle u_x u_y \rangle$, where $\langle u_x u_y \rangle$ is the relevant component of the Reynolds tensor $\langle u_\alpha u_\beta \rangle$. The square root of its trace is the fluctuation intensity $q_{\rm rms} = \sqrt{\langle u_\alpha u_\alpha \rangle}$. The dissipation rate is $\epsilon = \langle 1/2\nu \left(\partial u_\alpha/\partial x_\beta + \partial u_\beta/\partial x_\alpha\right)\left(\partial u_\alpha/\partial x_\beta + \partial u_\beta/\partial x_\alpha\right)\rangle$, where $\nu$ is the kinematic viscosity of the fluid. Kolmogorov dissipative scale is given by $\eta = \left(\nu^3/\epsilon\right)^{1/4}$ and the Taylor microscale is defined via $\epsilon = 5\nu q_{\rm rms}^2/\lambda^2$. $\mathrm{Re}_\lambda = \lambda q_{\rm rms}/\nu$ is the relevant form of the Reynolds number.

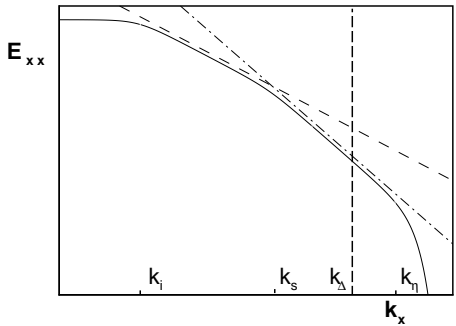

FIG. 2: Sketch of the spectrum $E_{xx}(k_x)$ (solid line). The dashed and dash-dotted lines correspond to $k_x^{-1}$ and to $k_x^{-5/3}$, respectively. The crossover is at $k_s = 2\pi/L_s$. The dissipative wavenumber is $k_\eta = 2\pi/\eta$ and the cut-off wavenumber, $k_\Delta = 2\pi/\Delta$, is within the inertial subrange, $k_s < k_x < k_\eta$.

The production range of shear flows - below $k_s$ in Fig. 2 - is less understood and, presumably, flow dependent. There a perturbation approach cannot work, since anisotropy is the prominent characteristics associated with the sustainment of turbulent fluctuations. Also in the production range, however, certain features seem to be universal. The spectrum displays a $k^{-1}$ power law - $E_{xx}(k_x) \propto k_x^{-1}$ - for $k_x < 2\pi/y$ near solid walls. Relative exponents - $\zeta_{rel}(n) = \zeta(n)/\zeta(3)$ [6] - show systematic differences with respect to classical results [7–9] and consistent deviations are found in direct numerical simulations (DNS), either in presence [10–12] or absence of solid boundaries [13]. All these indications support the conjecture of a universal regime. However a direct evidence, e.g. in terms of scalings in physical separation, has been missing so far.

Several reasons may have prevented the identification of a shear-dominated scaling regime, $S_n(r) \propto r^{\zeta_S(n)}$. On the one hand, limitations on the Reynolds number of present-day DNS hinder the direct observation of scaling laws. On the other hand, experiments cannot provide the spatial information to distinguish the contributions of different SO(3) sectors and this, as we shall see, may lead to an ambiguous physical interpretation.

Purpose of the present letter is to assess the existence of shear-dominated scaling laws in the simplest conditions of a homogeneous shear flow, see Fig. 1. The combined use of experimental and numerical data will enable us to extract the exponents for the isotropic sector to a sufficient degree of accuracy to claim that their values definitely differ from those of the inertial subrange.

We address different data sets. The experimental ones are described in [14]. The others are provided by highly resolved large eddy simulations (LES) at three different shear intensity. The LES approach by achieving sufficiently large Reynolds numbers allows for scaling laws in terms of separation to emerge.

**II. Large Eddy Simulation.** In LES the coarse grained field $\bar{u}_\alpha(x_\beta, t)$ is related to the fine grained field $u_\alpha(x_\beta, t)$ via a spatial filter $\mathcal{G}$ with cut-off length $\Delta$ [15],

$$\bar{u}_\alpha(x_\beta, t) = \int G_\Delta(x_\beta - \tilde{x}_\beta)\, u_\alpha(\tilde{x}_\beta, t) d^3\tilde{x} \; . \quad (1)$$

The flow has the mean profile sketched in Fig. 1 and periodic initial conditions on the fluctuating field. The equation for the coarse grained solenoidal field is

$$\frac{\partial \bar{u}_\alpha}{\partial t} = \epsilon_{\alpha\beta\gamma}\overline{u_\beta \zeta_\gamma} - \partial_\alpha \bar{\pi} + \nu\, \partial_{\beta\beta} \bar{u}_\alpha - S \bar{u}_y \delta_{\alpha 1} - \frac{\partial \overline{u_\alpha U}}{\partial x} \; , \quad (2)$$

where $\pi$ is the sum of pressure and kinetic energy density and the permutation symbol $\epsilon_{\alpha\beta\gamma}$ performs the cross product between velocity and its curl $\zeta_\alpha$.

Equation (2) alone is insufficient to determine $\bar{u}_\alpha$ and a suitable closure is needed. The approximate deconvolution method (ADM [16]) uses an approximate inversion of the filter to reconstruct $u_\alpha$ from $\bar{u}_\alpha$ and evaluate the unclosed terms $\epsilon_{\alpha\beta\gamma}\overline{u_\beta \zeta_\gamma}$ and $\overline{u_\alpha U}$.

The simulation follows the standard, state-of-the-art numerical procedure for homogeneous turbulence subject to uniform shear [17, 18]. Here the filter cut-off $\Delta$ is always much smaller than $L_s$ to have the solution virtually unaffected by the model in the production range. The resolution is $N_x \times N_y \times N_z = 192 \times 216 \times 96$, with $St_{max} \simeq 500$ and $Dt(\epsilon/\eta^2)^{1/3} \simeq .03$.

**III. Scalings in the shear dominated range.**

The homogeneous shear flow is characterized by two independent parameters, $S^* = Sq_{rms}^2/\epsilon$ and $Re_\lambda = q_{rms}\lambda/\nu$. Longitudinal structure functions

$$S_n(r_x) = \langle (u_x(x + r_x, y, z) - u_x(x, y, z))^n \rangle \quad (3)$$

from simulations and experiments are compared in Fig. 3.

In the two cases - experiments and LES - the structure functions agree very well, not only in shape but also in magnitude. A scaling behavior in terms of separation is clearly inferred from the plots and confirmed by the inset. Experiments and numerics provide identical exponents. In comparison, the exponents of isotropic turbulence are considerably different, see the figure caption.

Longitudinal structure functions follow from the superposition of all the SO(3) sectors [3],

$$S_n(r_x) = \sum_{j=0}^{\infty} \sum_{m=-j}^{j} S_{jm}^{(n)}(r)\, Y_{jm}(0, 0)\,, \quad (4)$$

where $Y_{jm}(\theta, \phi)$ are the spherical harmonics in terms of polar, $\theta$, and azimuthal, $\phi$, angles. Here $j$ denotes the sector, with $j = 0$ the isotropic contribution independent of the orientation of the radial vector, and $S_{jm}^{(n)}(r)$, $-j \leq m \leq j$, the $2j+1$ components of the structure function in the $j^{\rm th}$ sector. In principle, scaling laws in the different sectors with j-dependent exponents, $S_{jm}^{(n)}(r) \propto r^{\zeta_j(n)}$, may lead to a blending of power laws, and to an effective logarithmic slope in the complete longitudinal structure functions. If this is the case, the effective slope would strongly depend on the relative intensity of the different sectors, resulting in a highly flow dependent feature. We will see below that the scaling laws emerge in a different way.

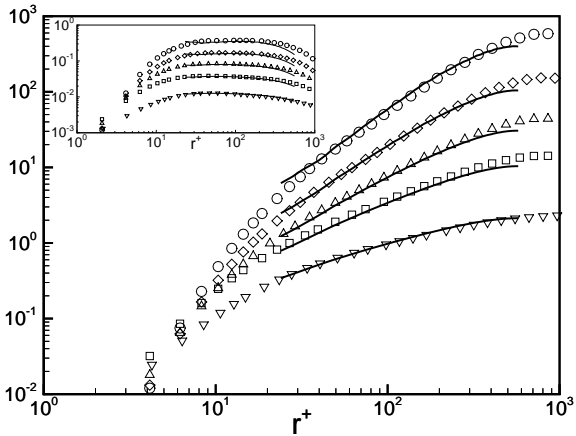

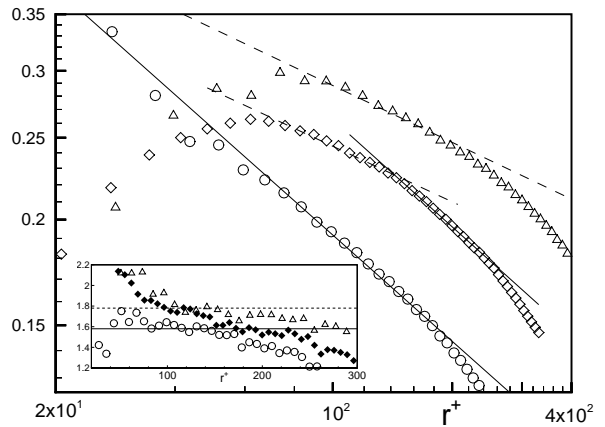

FIG. 3: Longitudinal structure functions vs separation, experiments (symbols) - $n = 2$ (triangles), $n = 3$ (squares), $n = 4$ (delta), $n = 5$ (diamonds) and $n = 6$ (circles) - and LES (solid line). Moments $n = 3, 4, 5, 6$ are multiplied by 2. Data normalized by $q_{\rm rms}$, $r^+ = r/\eta$. In the experiment $\Pi \simeq \epsilon \simeq .85 {\rm m}^2{\rm s}^{-3}$, $S \simeq 19.{\rm s}^{-1}$, $q_{\rm rms} \simeq \sqrt{u_{\rm rms}^2 + 2v_{\rm rms}^2} \simeq .6 {\rm ms}^{-1}$, $S^* \simeq 8$ and $Re_\lambda \simeq 220$. LES at $S^* = 7$ and $Re_\lambda = 150$, $\Delta^+ = \Delta/\eta = 17$, $L_{\rm s}^+ = L_{\rm s}/\eta = 70$. Computational box: $L_{\rm x} = 4\pi$, $L_{\rm y} = 2\pi$, $L_{\rm z} = 2\pi$, with $L_{\rm x}^+ = 1250$. Concerning the integral scale $L = q_{\rm rms}^3/\epsilon$, $L^+ = L/\eta = 1400$. In the inset, the same data in compensated form, $S_n/r^{\zeta_{\rm s}(n)}$, same symbols: $\zeta_{\rm s}(n) = 0.72, 1.00, 1.23, 1.42, 1.58$, for $n = 2, 3, 4, 5, 6$, respectively (For isotropic flows $\zeta(n) = 0.69, 1.00, 1.28, 1.54, 1.78$).

FIG. 4: Isotropic component of the sixth order longitudinal structure function normalized by its dimensional scaling, $S_{00}^{(6)}/r^2$. High shear case: circles, parameters defined in the caption of Fig. 3. Low shear case: triangles, $S^* = 2.2$, $Re_\lambda = 160$, $L_{\rm s}^+ = 430$, $L^+ = 1325$, $L_{\rm x}^+ = 1630$, $\Delta^+ = \Delta/\eta = 20$. Intermediate shear case: diamonds, $S^* = 5.4$, $Re_\lambda = 150$, $L_{\rm s}^+ = 110$, $L^+ = 1320$, $L_{\rm x}^+ = 1320$, $\Delta^+ = \Delta/\eta = 17$. The slope of the solid lines is $-.42$, corresponding to $\zeta_{\rm s}(6) = 1.58\pm.08$. For the dashed lines the slope is $-.22$, i.e. $\zeta(6) = 1.78\pm.08$. In the inset, the local slope (diamonds filled for better readability).

In order to ascertain the physical origin of the scalings reported in Fig. 3 it is appropriate to extract the dominant contributions from the general SO(3) decomposition. The interest is naturally focused on the isotropic sector, $j = 0$. Unfortunately experimental data can hardly be used to cleanly extract the pure isotropic component, $S_{00}^{(n)}(r)$, due to the lack of spatial information. However this is easily done on the basis of numerical data, considering that $Y_{00}(\theta,\phi) \equiv 1/\sqrt{4\pi}$, so that

$$S_{00}^{(n)}(r) = \frac{1}{\sqrt{4\pi}} \int_0^\pi \int_0^{2\pi} S_n(r_{\rm x}, r_{\rm y}, r_{\rm z}) \sin(\theta)\, d\theta d\phi \ . \quad (5)$$

Results for $n = 6$, where the difference between the exponent of the longitudinal structure function - $\zeta_{\rm s}(6) = 1.58$ - and that of isotropic turbulence - $\zeta(6) = 1.78$ - is particularly large, are reported in Fig. 4. Here the isotropic contribution $S_{00}^{(6)}(r)$ is plotted for three different shear intensity $S^*$. The first one, the high shear case (circles), is the same we have already discussed. It is clear from the plot that 1.58 (solid line) is the appropriate exponent to achieve compensation. It is also clear from the figure that the classical isotropic value of 1.78 (dashed lines) is unsuitable to fit the data. We conclude that a pure scaling law with an exponent considerably smaller than the isotropic one characterizes the isotropic sector of this shear dominated flow. This exponent is able to fit also the longitudinal structure functions, as already shown in Fig. 3. The SO(3) decomposition allows here to exclude contamination effects from the anisotropic sectors.

The difference between 1.58 and 1.78 represents the shear-induced alteration of the exponent in the isotropic sector of the sixth order structure function. Contrary to expectation, this effect is not a flow dependent superposition of universal exponents of all sectors, but the modification of the exponent of the isotropic sector itself.

As we may expect from the previous analysis, below the shear scale we recover the isotropic exponents. Going below $L_{\rm s}$ still remaining sufficiently apart from dissipative effects requires some care in the homogeneous shear flow. To this purpose we have selected the low shear case defined in Fig. 4 (triangles) [19, 20]. The plots in the figure are self-explanatory: below the shear scale we actually reproduce the results of isotropic turbulence. Ideally, when the shear scale lays in the middle of the scaling range, one should observe the simultaneous presence of the two scaling behaviors. This is what happens indeed in the third case (diamonds) where the isotropic component of the sixth order structure function scales with $\zeta(6) = 1.78$ (dashed line) at small scales and with $\zeta_s(6) = 1.58$ at large scales. With acceptable precision, the cross-over is found at $r = L_s$.

**IV. Final comments.** In conclusion, we have produced scaling laws in physical separation for the shear-dominated range of the homogeneous shear flow. The exponents are lower than those of the classical inertial subrange consistently with the larger intermittency of the production scales.

It should be mentioned that the fitting range must be carefully selected for accurate extraction of the pure exponents. Different problems may emerge, in general, when addressing data pertaining to shear flows. Typically, in experiments the shear scale is not always known with sufficient precision due to uncertainty in the evaluation of $\epsilon$. Moreover, in wall bounded flows the near-wall region can be resolved only at small Reynolds number.

At this point it is natural to ask ourselves the following questions. Are the scaling exponents universal for any kind of shear flow? In other words, may we transfer what we have found in the homogeneous shear flow to all other shear-dominated flow conditions? The answer may be given by an indirect approach. One of the present non-trivial results is that $S_3(r_x)$ scales linearly with separation in the production range, see $\zeta_s(3)$ in the caption of Fig. 3. Hence, the relative exponents extracted from shear flow data can be used to infer $\zeta_s(n) = \zeta_{\rm rel}(n)$ [6]. Actually, all the data available to us - channel flows, boundary layers at different distance from the wall, homogeneous shear flows and presumably even jets - yield the exponents we have produced here. In other words, the production range of shear flows seems to be characterized by the universal set of scaling exponents given in Fig. 3. From the point of view of applications such universality is instrumental for developing proper closures for LES of wall bounded flows.

---